\documentclass[12pt]{article}

\usepackage{scicite}

\usepackage{times}
\usepackage{gensymb}
\usepackage{graphicx}

\DeclareUnicodeCharacter{2212}{-}

\newenvironment{sciabstract}{%
\begin{quote} \bf}
{\end{quote}}

\newcounter{lastnote}

\title{Measurements of CNT Forest Self-Assembly from In-situ ESEM Synthesis}

\author
{Ramakrishna Surya,$^{1}$ Gordon L. Koerner,$^{1}$ Taher Hajilounezhad,$^{1}$\\ Kaveh Safavigerdin,$^{2}$ Prasad Calyam,$^{2}$ Filiz Bunyak,$^{2}$ \\Kannappan Palaniappan,$^{2}$ Matthew R Maschmann,$^{1,3\ast}$\\
\\
\normalsize{$^{1}$Department of Mechanical and Aerospace Engineering, University of Missouri,}\\
\normalsize{Columbia, MO, 65211, USA}\\
\normalsize{$^{2}$Department of Electrical Engineering and Computer Science, University of Missouri,}\\
\normalsize{Columbia, MO, 65211, USA}\\
\normalsize{$^{3}$MU Materials Science and Engineering Institute, University of Missouri,}\\
\normalsize{Columbia, MO, 65211, USA}\\
\\
\normalsize{$^\ast$To whom correspondence should be addressed; E-mail:  maschmannm@missouri.edu.}
}

\date{}

\begin{document} 


\baselineskip24pt


\maketitle


\begin{sciabstract}
Understanding and controlling the dynamic self-assembly mechanisms of carbon nanotube (CNT) forests is necessary to advance their technological impact. Here, in-situ environmental scanning electron microscope (ESEM) chemical vapor deposition (CVD) synthesis observes the real-time nucleation, assembly, delamination, and self-termination of dense  ($>$ 10$^9$ CNT/cm$^2$), tall ($>$ 100 $\mu$m) CNT forests in real time. Forest synthesis is continuously observed from nucleation to self-termination. Assembly forces generated near the substrate detach CNTs from the substrate, which simulation suggests requires approximately 10 nN of tensile force. Delamination initiates at both the CNT-catalyst and the catalyst-substrate interfaces, indicating multiple delamination mechanism. Digital image correlation applied to SEM image sequences measures time-invariant strain within growing forests, indicating that forests grow as rigid bodies after liftoff. The Meta CoTracker algorithm measured CNT growth rates reduce from 50 nm/sec to full termination over 150 seconds. This work provides a robust strategy to observe and measure CVD material synthesis in-situ using ESEM. The method is uniquely suited to observe population-based phenomena at both nanometer spatial resolution and at a highly scalable field of view.
\end{sciabstract}

% In setting up this template for *Science* papers, we've used both
% the \section* command and the \paragraph* command for topical
% divisions.  Which you use will of course depend on the type of paper
% you're writing.  Review Articles tend to have displayed headings, for
% which \section* is more appropriate; Research Articles, when they have
% formal topical divisions at all, tend to signal them with bold text
% that runs into the paragraph, for which \paragraph* is the right
% choice.  Either way, use the asterisk (*) modifier, as shown, to

\section{Introduction}

CNT forests are electrically and thermally conductive, soft, durable, and are nearly perfect optical absorbers. The blend of mechanical, thermal, and electrical properties are advantageous in applications ranging from multi-functional structural materials \cite{garcia,wicks}, flexible sensors \cite{ehlert,maschmannForceSens}, electrochemical energy storage \cite{lee2010high,zhu2021embedding,carter2017high}, electronics \cite{park}, and conductive interface materials \cite{xu,park}. The physical properties of CNT forests are significantly diminished relative to those scaled from individual CNTs. For example, the elastic modulus of an individual CNT may exceed 1 TPa, whereas the modulus of CNT forests is on the order of 1-100 MPa \cite{zhang, yaglioglu2012wide}. The degradation in properties originates from the highly tortuous and disorganized morphology of CNTs within a forest. The mechanistic processes guiding CNT forest self-assembly, and thus their material properties,  remain uncertain \cite{maschmannForceSens,zhang,zbib,wheeler,tawfik,Davis_Hierarchical,hutchens,cao,hart,pathak}, and insights into the process-structure-property relationships of CNT forests are limited. 

Measuring the dynamic morphology evolution of CNT forests during synthesis poses several challenges because of the small observation length scale (sub-$\mu$m), high synthesis temperatures ($>$ 600 $\degree$C), and the need for precise control of the chemical environment. Digital video acquisition of a growing CNT forest may be achieved by placing a digital camera in-line with the process tube of a conventional chemical vapor deposition system \cite{roboFurnace}. Such a system can measure CNT forest height with a resolution on the order of 20 $\mu$m. Laser reflectivity from CNT forests grown in cold-walled chemical vapor deposition (CVD) chambers have measured the time-resolved height of growing CNT forests with a resolution on the order of 10 nm \cite{laserReflect}, but the technique is limited to height measurements alone. In-situ small angle X-ray scattering (SAXS) can measure population-based metrics including diameter, alignment, and areal density \cite{Meshot_HighSpeed, PopulationGrowthSAXSBedewy}, but not individual CNT-CNT interactions \cite{abruptTermination}. Assuming an X-ray beam diameter of 100 $\mu$m \cite{Meshot_HighSpeed}, SAXS measurements sample approximately 10$^7$ CNTs when interrogating a 1 x 1 cm CNT forest with an areal density of 10$^9$ CNT/cm$^2$. SAXS measurements have contributed greatly to the understanding of population-level forest morphology evolution, including a decreasing CNT population density and CNT alignment with increasing growth time. The mechanisms promoting these behaviors, however, are inaccessible using SAXS alone.  

\begin{figure}[ht]
  \includegraphics[width=85mm]{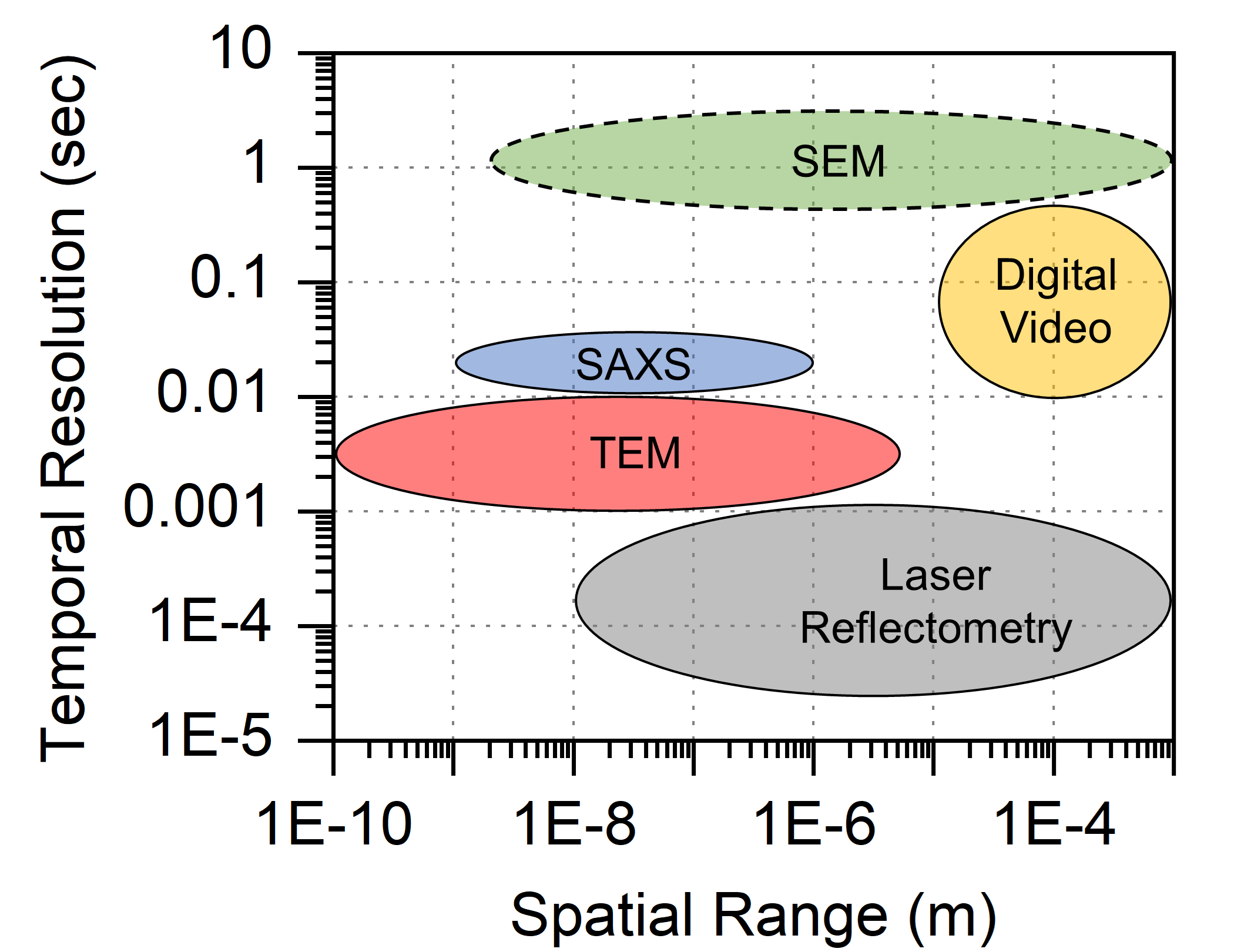}
  \centering
  \caption{Schematic mapping of various in-situ CNT forest synthesis measurement techniques. The applicable spatial domain of a technique is represented on the horizontal axis. The minimal temporal resolution is represented on the vertical axis. Of these techniques, only TEM and SEM can locally resolve individual CNTs and their dynamic behavior. Laser reflectivity and digital video measure the height of a CNT forest with time. SAXS resolves population-based distributions of CNT diameter, density, and alignment.}
  \label{fig:fig1}
\end{figure}

In-situ transmission electron microscope (TEM) synthesis methods, by contrast, directly capture real-time CNT synthesis imagery for both isolated CNTs and small CNT populations \cite{pattinson, balakrishnan, yasuda, bedewy, dee}. TEM enables sub-nanometer imaging resolution and can acquire images on the order of 100 - 1,000 frames per second. In-situ TEM synthesis is a cold-walled CVD process, wherein the observation region is comprised of an electron-transparent, thin-film heater. Reaction gases are supplied to the heated zone either through direct vapor delivery within an environmental TEM (ETEM) chamber \cite{pattinson, bedewy} or within a hermetically sealed, electron-transparent chamber established within the sample holder \cite{huang, SWCNT_Pt, InSitu_AmbientPressure}. Most in-situ TEM CNT syntheses have produced sparse and disorganized CNTs rather than dense and vertically oriented CNT forests. Such studies typically examined catalyst particle behavior during active CNT synthesis\cite{huang,bedewy}. Example in-situ TEM synthesis results include the dynamic motion of CNT catalyst particles, which can become partially incorporated into the inner diameter of CNTs during synthesis \cite{huang}. Other studies have demonstrated the mobility of iron catalyst nanoparticles on the surface of CNTs during CNT forest synthesis \cite{JeongTEM2016}. While TEM provides unparalleled image resolution of active catalyst nanoparticles and emerging CNTs, its field of view is limited, thereby restricting the observation of CNT forest population dynamics. For three-dimensional growth of CNT forests, TEM produces projected images through the thickness of a forest, obscuring depth perception and making image interpretation of high-density CNT forests challenging \cite{balakrishnan}. A schematic representation of minimum temporal resolution and total spatial resolution of existing techniques is provided in Figure 1.  

Here, we investigate the real-time in-situ CNT forest synthesis and self-assembly using  environmental scanning electron microscope (ESEM) technique on thin-film heating substrates similar to those used in TEM synthesis. ESEM observations capture CNT forest assembly across length scales ranging from individual CNTs ($<$ 10 nm) to global behaviors of growing CNT forest microstructures ($>$ 100 $\mu$m). We observed the CNT forest assembly process at the edge of well defined CNT forest micropillars to acquire unobstructed views of the relevant assembly processes. Image acquisition was acquired at one frame per two seconds at spatial resolutions comparable to conventional SEM imaging. By using a magnification of 50 - 100 kx, we observed critical processes at the level of individual CNTs, including nucleation, initial CNT self-organization, variations in CNT orientation over time, CNT detachment from the growth substrate, and self-termination. Global CNT forest measurements of strain and growth rate were facilitated by digital image correlation (DIC) software, CNT area density was quantified by hand-crafted analysis methods, and the growth rate of individual CNTs during self-termination was measured using the Meta CoTracker algorithm. Adventitious carbon deposition, which is a common product of SEM imaging, played an unexpected vital role in promoting localized catalyst reduction and the formation of tall CNT forest micropillars. The CNT forests synthesized by in-situ ESEM synthesis closely resemble the areal and density of CNT forests synthesized by atmospheric pressure hot-walled CVD synthesis.

\section{Results and Discussions}
\subsection{Catalyst Reduction, CNT Nucleation, and Early CNT Growth}
Prior to CNT synthesis, a catalyst iron (Fe) thin film must undergo dewetting to transform into discrete catalytic nanoparticles. In our experiments, catalyst reduction and dewetting was spatially controlled by patterned adventitious carbon deposited by SEM electron beam rastering prior to substrate heating. Carbonaceous species, whether in gaseous or solid form, facilitate the efficient reduction of the iron thin film at low pressure, promoting the growth of dense CNT forests \cite{dee, carpena}. The carbon was deposited by repeatedly scanning the electron beam over the isolated regions targeted for patterned CNT forest growth for up to 20 minutes. These depositions were clearly visible in SEM imagery before the synthesis process, as illustrated in Figure \ref{fig:fig1}a. The array of dark squares represent varying levels of carbon deposition, as controlled by the duration of electron beam scanning. Upon heating, iron nanoparticles formed and appeared as a field of bright particles within the carbon deposition region, as depicted in Figure \ref{fig:fig1}b. A more detailed video of the dynamic formation of iron nanoparticles is available in the supplementary information video S1. The catalyst nanoparticles shown in \ref{fig:fig1}b were formed at a temperature of  550 \degree C in a high vacuum environment without the intent of CNT synthesis. Note that particle formation occurred below the CNT synthesis temperature of 625 \degree C. Dense CNT forest micropillars grew from the carbon deposition regions in Figure \ref{fig:fig1}a, as shown in Figure \ref{fig:fig1}c. CNT forests with heights exceeding 100 $\mu$m occurred within three of the four regions of carbon deposition, representing 10, 15, and 20 minutes of continuous electron beam carbon deposition. Only sparse CNT synthesis was observed in the region corresponding with the lowest carbon dose corresponding with 5 minutes of deposition. The nearly binary variation in CNT forest height indicates that a minimum dose of carbon is needed to support robust catalyst reduction and subsequent CNT forest synthesis. Numerous unsuccessful syntheses occurred in the absence of adventitious carbon, characterized by sparse or no observable CNT growth.
  
  \begin{figure}[ht]
  \includegraphics[width=165mm]{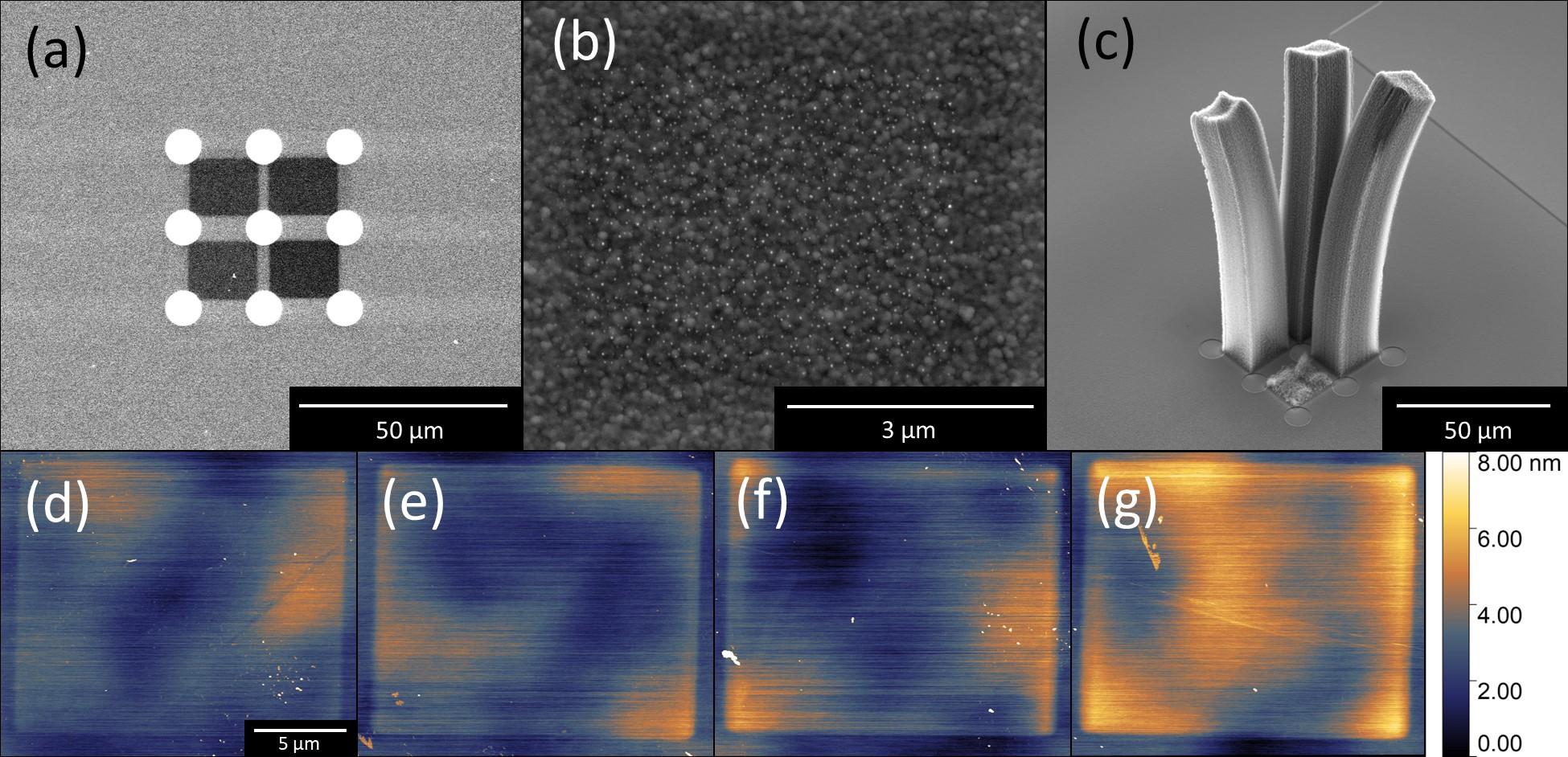} 
  \centering
  \caption{Adventitious carbon deposition enabled patterned CNT synthesis. \textbf{a} SEM image of carbon deposition on a Protochips substrate using deposition times of 5, 10, 15, and 20 minutes. \textbf{b} SEM image of preferential reduction of an iron catalyst film when exposed to 550 $\degree$C in high vacuum. \textbf{c} SEM image of the CNT pillars from the substrate shown in \textbf{a}. \textbf{d-g} AFM topology of adventitious carbon deposited on a silicon wafer using electron beam exposure times from 5 - 20 minutes, respectively.}
  \label{fig:fig1}
\end{figure}

 Atomic force microscope (AFM) topography images of adventitious carbon deposited on silicon wafers show the relationship between carbon deposition and electron beam scan time. The deposition regions shown in Figure\ref{fig:fig1}(d-g) employed the same SEM carbon deposition parameters used in Figure\ref{fig:fig1}a. The coverage area and thickness of the deposition increased with longer electron beam deposition times. The greatest carbon deposition often occurs near the boundaries of the scanned areas where electron beam scanning terminates, resulting in a non-uniform carbon topology within the deposition zone. It is unclear how the carbon migrates upon heating and catalyst dewetting. Nevertheless, it is evident that the patterning of carbon enabled well-defined activity of iron catalyst nanoparticles, which in turn enabled in-situ observation of well-defined CNT forest edges.

\subsection{CNT Forest Assembly}
The well-defined vertical faces of CNT forest pillars offered an ideal platform for observing the growth and assembly of CNTs at all magnifications. Individual CNT behaviors were readily discernible between 50,000-100,000 x magnification. The analysis of bulk, monolithic micro-pillar behavior was conducted at approximately 10,000 x magnification. High magnification examination demonstrated that the CNTs followed the base-growth mechanism, wherein the catalyst for CNT synthesis remained on the substrate during the growth process. This observation aligns with the findings of previous studies that employed similar catalyst systems.

\begin{figure}[ht]
  \includegraphics[width=165mm]{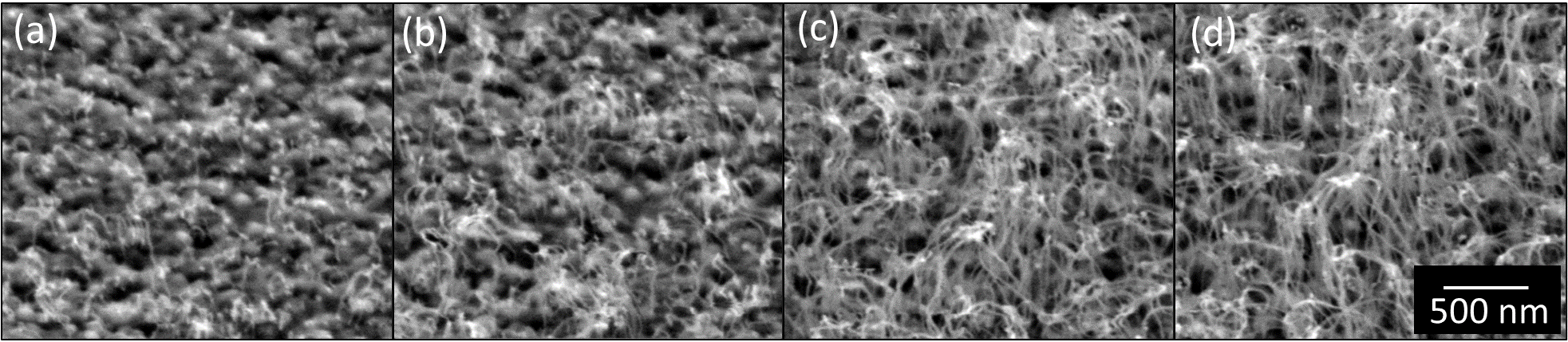}
  \centering
  \caption{Early stages of CNT forest self-assembly. An in-situ ESEM synthesis image sequence shows CNT growth at \textbf{a} t = 2, \textbf{b} t = 4 seconds, \textbf{c} t = 6 seconds and \textbf{d} t = 8 seconds after nucleation. \textbf{a} CNTs initially form arcs along their length, with the base and tip contacting the substrate. Growth is independent of neighbors. \textbf{b} Upon achieving approximately 1$\mu$m of growth, neighboring CNTs cling together via van der Waals force, forming \textbf{c} localized colonies of entangled and vertically oriented CNT tufts. Each tuft grows in height and breadth \textbf{d}, eventually contacting neighboring colonies and forming a continuous CNT forest. }
  \label{fig:LifeSequence}
\end{figure}

The catalyst reduction and early CNT forest assembly obtained in a 10 Pa C$_2$H$_2$ environment is shown in Figure \ref{fig:LifeSequence}.  After 28 seconds of heating, bright catalyst nanoparticles formed at a temperature of 425$\degree$C, signaling the dewetting of iron nanoparticles. Approximately 10 seconds later (575$\degree$C), CNTs nucleated from a fraction the visible catalyst particles. Early CNTs exhibited mainly an arched morphology, where the CNT tips and bases were in contact with the growth substrate. As the CNTs lengthened, neighboring CNTs contacted each other due to crowding, forming van der Waals bonds between contacting CNTs. Contacting CNTs formed small assemblies that lifted away from the growth substrate. Over time and with increased CNT growth, isolated assemblies merged together to form a fully connected CNT forest. The early CNT-CNT assembly appeared similar to that observed by in-situ TEM CNT forest synthesis, where the density of CNT-CNT contacts significantly increased with time \cite{balakrishnan}.  A video of the sequence in Figure \ref{fig:LifeSequence} can be found in supplemental information video S2.

After fully connected CNT forests formed, CNT-CNT and CNT-substrate interactions were observed at the edge of the micropillars. The mechanical degrees of freedom at the CNT base may be inferred by the change in orientation angle between a CNT and the substrate in response to mechanical loading. The deformation of CNTs allowed by this boundary condition plays a critical role in CNT forest morphology evolution, as it determines rotational mobility. A time-invariant orientation angle at the substrate would infer a fixed-end boundary condition that disallows CNT rotation. A time-varying angle would suggest that CNTs and/or catalyst particles can rotate or reconfigure in response to dynamic loads transmitted to the catalyst particles. 
  
\begin{figure}[ht]
  \includegraphics[width=85mm]{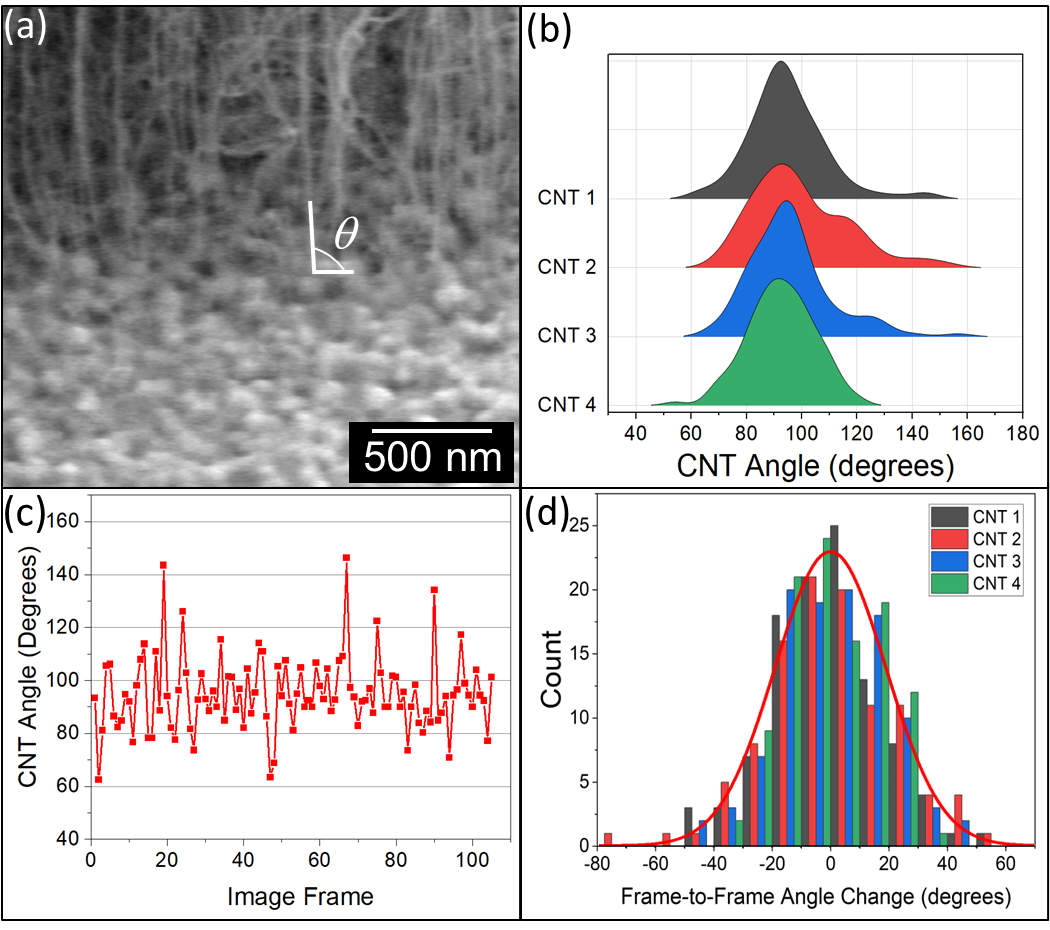}
  \centering
  \caption{Measuring the dynamic orientation angle of CNTs relative to the growth substrate.  \textbf{a} Example measurement scheme, as measured from in-situ ESEM image sequences. \textbf{b} Time-averaged histograms of 4 CNTs obtained over 105 sequential images (210 seconds). \textbf{c} A representative time sequence of CNT orientation angle as a function of time. \textbf{d} A histogram of frame-to-frame angle variation for four CNTs and an overlaid Gaussian fit to the population data.}
  \label{fig:figure3}
\end{figure}

We measured the projected rotation of four representative CNTs over 105 consecutive SEM imaging frames, corresponding to an observation of 210 seconds (Figure \ref{fig:figure3}b). The angle data demonstrates that CNTs are predominantly oriented at an angle of approximately 90 degrees, perpendicular to the growth substrate. Notably, CNTs labeled as CNT 1 and CNT 4 exhibit a nearly Gaussian angular distribution centered at 90$\degree$, while CNT 2 and CNT 3 show secondary peaks, indicating a secondary preferred orientation. The second orientation arose from repeated interactions between the measured CNT and a neighboring CNT. A time-resolved plot of CNT angle (Figure \ref{fig:figure3}c), demonstrates that the orientation angle of CNTs continuously vary, indicative of pinned mechanical boundary condition that permits rotation without translation. The rate of change in orientation angle between consecutive frames is depicted in Figure \ref{fig:figure3}d. A Gaussian fit to the cumulative frame-to-frame angle variation data generated by the four CNTs yields a mean angle change of 0.05$\degree$ per frame and a standard deviation of 19.0$\degree$ per frame. The average near zero degrees represents that the CNT oscillates at equal rates to the left and right relative to the observation frame. Occasional significant and abrupt changes in angle suggest that rapidly varying loads are transmitted to the CNT catalyst. The observed rotation may be accommodated by the CNT and catalyst system through CNT catalyst rotation about the substrate, catalyst reconfiguration, generation of CNT defects, CNT kinking, or rotation of the CNT about a fixed catalyst nanoparticle.

%\FloatBarrier

\subsection{Pillar Kinematics}
\begin{figure}[ht]
  \includegraphics[width=125mm]{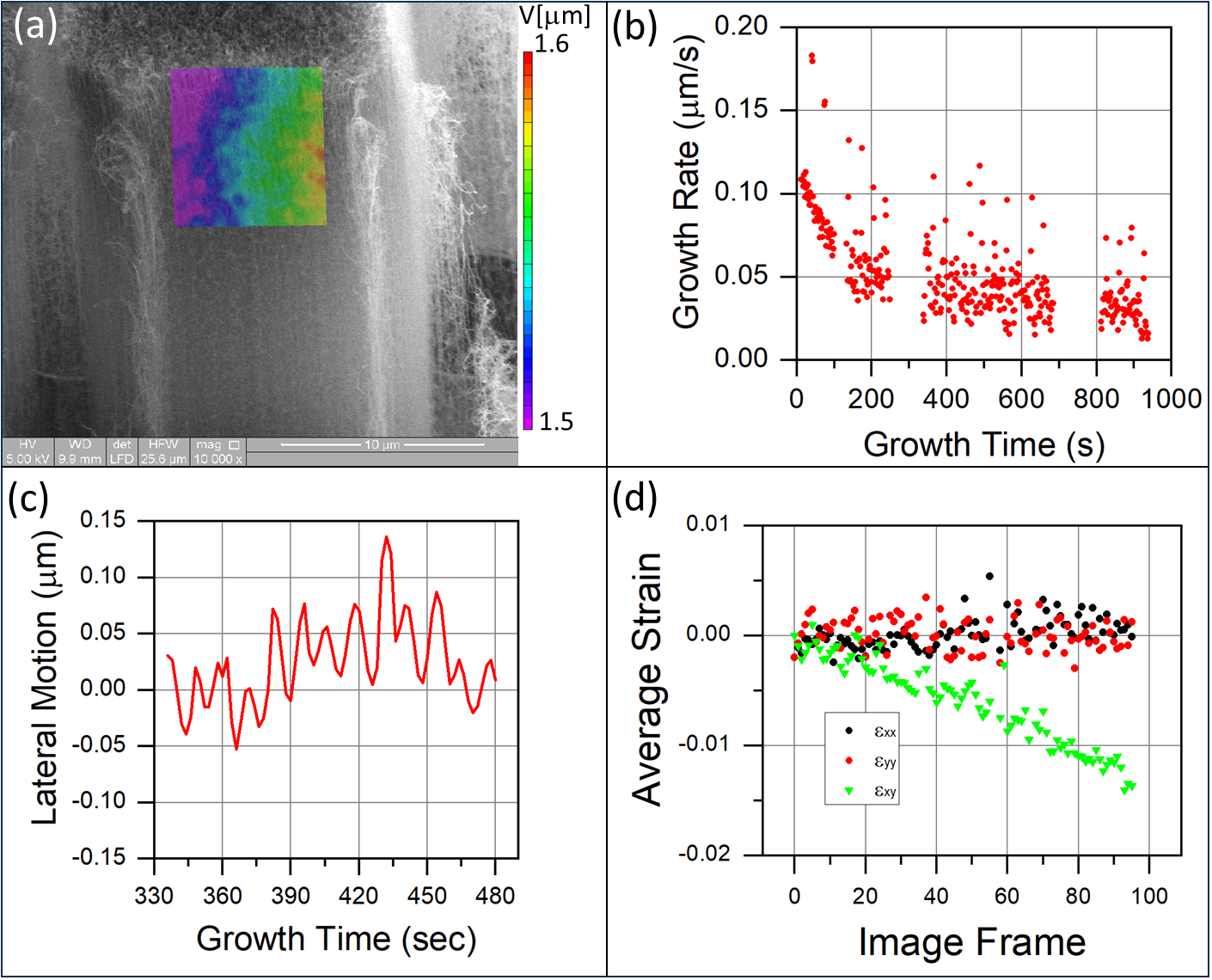}
  \centering
 \caption{Digital image correlation (DIC) analysis of in-situ ESEM CNT forest synthesis. \textbf{a} An in-situ CNT image overlaid with a DIC evaluation box measuring vertical displacement, V. \textbf{b} The average CNT forest vertical growth rate of the CNT forest pillar in \textbf{a} as a function of time obtained by averaging the vertical displacement within DIC evaluation boxes. \textbf{c} DIC measured lateral motion of the growing CNT forest pillar shown in \textbf{a}. \textbf{d} DIC measured lateral ($\epsilon_{x_x}$), vertical ($\epsilon_{y_y}$), and shear strain ($\epsilon_{x_y}$) for the CNT pillar shown in \textbf{a}. The constant $\epsilon_{x_x}$ and $\epsilon_{y_y}$ indicates that the pillar moves as a monolithic rigid body with invariant internal strain. Linearly decreasing shear strain is an indication that the CNT pillar is slowly growing at a tilted angle.}
  \label{fig:DIC}
\end{figure}

After lifting from the growth substrate, CNT forest micropillar monoliths grow away from the substrate. Digital image correlation (DIC) software was used to measure the collective CNT forest pillar growth rate, lateral translation, and internal strain. DIC algorithms evaluate pixel-by-pixel motion of a surface across an image sequence. Whereas a high-contrast speckle pattern is typically applied to surfaces to facilitate DIC measurements, the high-contrast pixel pattern generated by CNT forest SEM imagery is sufficient  \cite{MaschmannDIC,MaschmannContinuum}. Figure \ref{fig:DIC}a shows a representative DIC evaluation frame imposed on a growing CNT pillar. The color contours in the DIC evaluation box represent vertical displacement contours. The pillar displayed in Figure \ref{fig:DIC}a was monitored for 630 image frames, corresponding to a total observation time of 21 minutes. The vertical displacement measured by DIC, displayed in Figure \ref{fig:DIC}b, exhibits an exponential growth rate decay with time. The initial growth rate of 110 nm/s was measured 14 seconds after nucleation, and it decreased to 40 nm/s towards the end of observation. The measurement was unable to measure the beginning 14 seconds of growth rate because finite time was required to acquire focus and reposition the area of interest because of sample drift. Previous CNT forest growth rate studies used laser displacement and cold-walled CVD with ethylene feedstock at atmospheric pressure and 1,010 - 1,120 $\degree$C growth temperatures. These measurements showed a rapidly increasing growth rate at the onset of CNT forest synthesis, followed by a steady decrease in growth rate, and abrupt self-termination \cite{abruptTermination}. Optical measurements of hot-walled CNT forest growth rates showed a similar trend \cite{roboFurnace}.  While it is difficult to directly compare quantitative growth rate results between these experiments because of drastically different synthesis conditions, a decreasing CNT growth rate after initial rapid growth is consistent with prior observation.

Lengthening CNT forest micropillars exhibited coordinated lateral oscillations, normal to the growth direction, in addition to their vertical growth. The DIC-measured oscillating behavior of a representative CNT forest pillar is exhibited in Figure \ref{fig:DIC}c, with a peak-to-peak lateral translation of up to 130 nm. The oscillations were driven by the fastest-growing CNTs in the population, as their excess lengthening was resisted from above and was accommodated by coordinated lateral motion just above the substrate. This behavior is likely responsible for the coordinated waviness frequently observed within CNT forests.

The internal CNT forest strain parallel to the growth direction, $\sigma$$_{yy}$, and normal to the growth direction, $\sigma$$_{xx}$, remained invariant at zero strain, as shown in Figure \ref{fig:DIC}d. The DIC evaluation window was first established just above the growth substrate as a CNT volume lifted free of the substrate. A consistent zero strain indicates that mechanical strain is constant within a CNT forest after it lifts from the substrate and that the CNT forest acts as a rigid body. It is important to note that the strain measurement is relative to the strain state at which the DIC software could first register a consistent image speckle pattern, which occurred at about 1 $\mu$m above the substrate.  Non-zero retained mechanical strain must exist within the volume resulting from the tortuous CNT morphology. The DIC measurement indicates that the strain within the forest is retained and does not relax as a given volume travels farther from the substrate. A steadily decreasing shear stress, $\tau$$_{xy}$, corresponded to the gradual bending of the pillar towards the left as it vertically grows vertically upward. A representative DIC evaluation sequence video is shown in supplemental information video S3.
%\FloatBarrier

\subsection{CNT Delamination}
\begin{figure}[ht]
  \includegraphics[width=85mm]{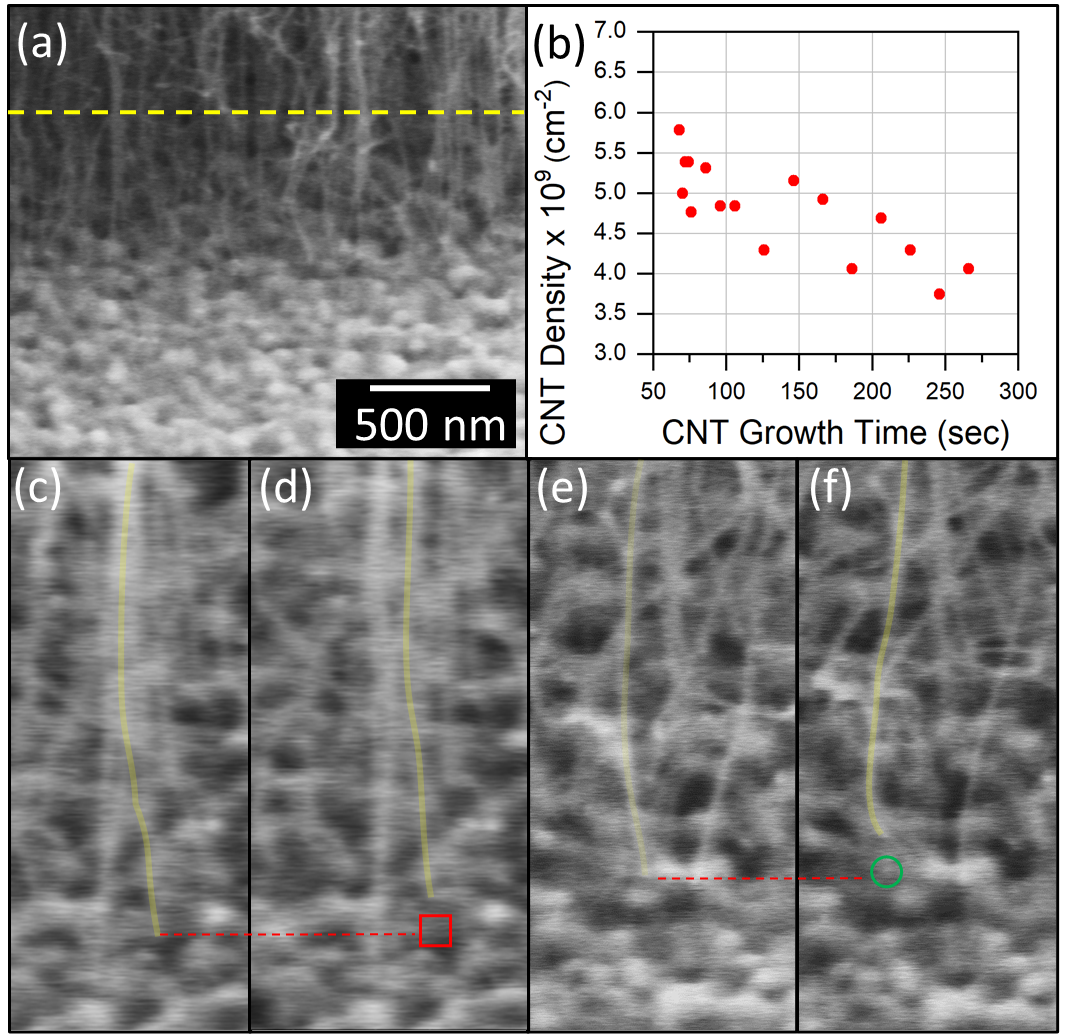}
  \centering
 \caption{Direct observation of CNT forest areal density decay. \textbf{a} The CNT areal density was measured by counting grayscale intensity peaks across a horizontal line overlaid to an in-situ ESEM image sequence. \textbf{b} The areal density decreased with synthesis time. Individual CNT detachment events, including \textbf{c-d} show CNT detachment at the catalyst-substrate interface. No catalyst particle is observed after CNT liftoff, as indicated by the square in \textbf{d}. \textbf{e-f} Before and after images showing the detachment of a CNT at the CNT-catalyst interface, with the substrate-adhered catalyst particle circled in \textbf{f}. The detached CNT is highlighted in yellow in both image sequences. These delamination events may also be observed in the Supplementary Information video S4.}
  \label{fig:DelamSequence}
\end{figure}

A persistent temporal decrease in CNT areal density was observed for all CNT syntheses. Previous in-situ SAXS measurements showed that areal density decreased by approximately half within the first 200 $\mu$m of growth of CNT forests grown at atmospheric pressure \cite{dee}. In-situ ESEM provides sufficient spatiotemporal resolution to observe both the individual CNT detachment events that SAXS can not discern and the CNT forest global density. Using ImageJ analysis software, a horizontal analysis line was projected across the field of view on image sequences, and the gray-scale pixel intensity profile was measured across the line (Figure \ref{fig:DelamSequence}a). Each peak in intensity profile represented a CNT or small CNT bundle intersecting the boundary. The intensity peaks were identified and counted using the Peak Analyzer tool in Origin 2022 software. The density analysis encompassed a 200-second evaluation window beginning 66 seconds after the initiation of CNT growth. The 66 second duration represents the time required to focus and center the observation area after initial thermal drift. The CNT density evolution is presented in Figure \ref{fig:DelamSequence}b. The horizontal field width of the image sequence was 2.56 $\mu$m, and the depth of field was estimated to be 1 $\mu$m. The number of CNTs decreased from 74 in the first frame (66 seconds after CNT growth initiation) to 52 CNTs after 200 seconds, or approximately 20 $\mu$m of CNT forest growth. These CNT counts correspond to an absolute CNT density ranging from approximately 6 x 10$^9$ to 4 x 10$^9$ CNT/cm$^2$, respectively, consistent in scale with previous SAXS measurements \cite{Meshot_HighSpeed, PopulationGrowthSAXSBedewy}. Although the SEM cannot differentiate between individual CNTs from small CNT bundles, it provides a reasonable and consistent estimate of CNT density as a function of time. 

Multiple CNTs were directly observed detaching from the growth substrate. Two such CNT detachment events are shown in Figure \ref{fig:DelamSequence}c-f. Some CNT delamination events appeared to leave a catalyst particle adhered to the substrate (Figure \ref{fig:DelamSequence}c-d), while others appeared to remove the catalyst particle from the substrate (Figure \ref{fig:DelamSequence}c-d). The presence or absence of a catalyst particle provides insight into the detachment mechanism and the relative energy of adhesion. Detachment may occur between the CNT-catalyst particle interface or the catalyst particle-substrate interface according the the adhesion strength in those two regions. Prior adhesion tests of CNT forests conducted at room temperature observed that delamination events of as-grown CNT forests occur preferentially at the CNT-catalyst \cite{WardleAdhesion} or a mix of CNT-catalyst and catalyst-substrate delamination \cite{brownDelam}. Though not explicitly discussed, particle-substrate detachment was observed in prior in-situ TEM studies \cite{balakrishnan} (see Supplementary Information). Our in-situ SEM observations indicate that a mix of both mechanisms are active at the elevated temperature of the synthesis environment. A video highlighting CNT detachment may be found in supplemental information video S4.

Utilizing a 3D mechanical finite element simulation, we quantified the forces conveyed to catalyst particles at the base of each CNT during CNT forest synthesis and self-assembly \cite{hajilounezhad2019evaluating, maschmann2015integrated}. To ascertain the upper limits of anticipated CNT delamination force, the simulation postulated continuous substrate adhesion and the absence of CNT delamination. In this computational framework, all CNTs within the forest were characterized by an outer radius of 10 nm and an inner radius of 5 nm. Consistent with in-situ observations, CNT-CNT contacts were persistent once established. The CNT population growth rate reflected log-normal distribution, with a population averaged growth rate of 60 nm per iteration and a standard deviation of 10\% of the mean. The simulation spanned a square region consisting of 100 x 100 CNTs at a density of 6 x 10$^9$ CNT/cm$^2$ over 200 time steps, yielding a final CNT forest height of approximately 10 $\mu$m. A density 6 x 10$^9$ CNT/cm$^2$ was selected as a CNT density to be consistent with our experimentally observed initial areal densities (Figure \ref{fig:DelamSequence}b). 

\begin{figure}[ht]
  \includegraphics[width=85mm]{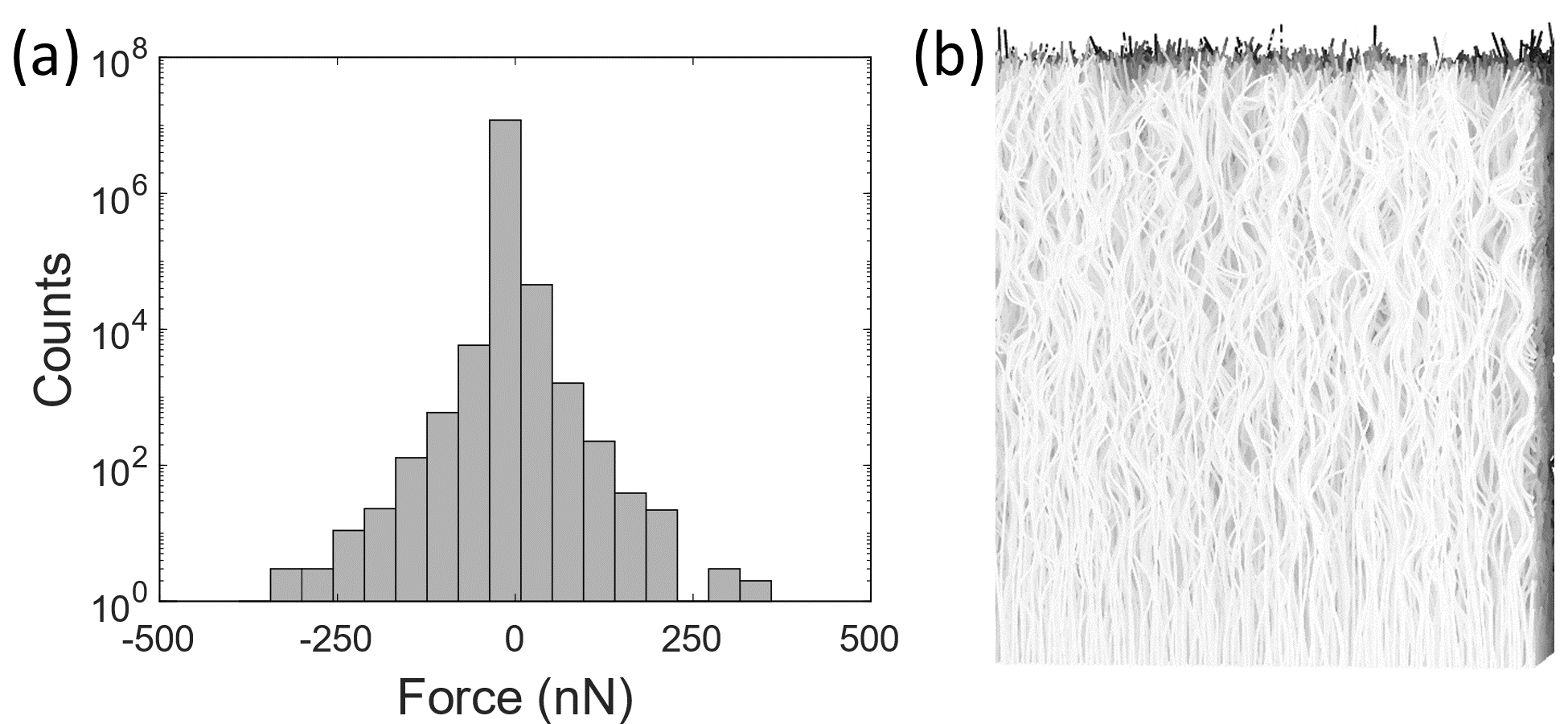}
  \centering
 \caption{A 3D mechanical finite element simulation evaluated the self-assembly force exerted at the CNT-substrate interface for 10,000 CNTs and approximately 10 $\mu$m of height. \textbf{a} A histogram of the CNT-substrate force spanned from -488 nN (compression) to 342 nN (tension) when CNT detachment was disallowed, with most recorded forces near zero load. \textbf{b} The resulting simulated CNT forest morphology resembles CVD-synthesized CNT forests.}
  \label{fig:Simulation}
\end{figure}

A histogram depicting the axial force transmitted to the base of each CNT for all time steps is presented in Figure \ref{fig:Simulation}a. Over the course of 200 simulated time steps, the maximum simulated force exerted between CNTs and the substrate ranged from +342 nN (tension) to -488 nN (compression). Previous work by Brown et. al \cite{brownDelam} established a critical tensile loading threshold of 12 nN/CNT to delaminate a CNT forest from its substrate at room temperature. Others have estimated the delamination force of between 1-3 nN per CNT \cite{WardleAdhesion}. The simulated forces generated during self-assembly exceed these critical values by at least an order of magnitude, suggesting that delamination is likely during the synthesis process. Using 12 nN as a critical tensile load to activate delamination, 1460 unique CNTs, corresponding to 14.6 \% of the CNT population, experienced sufficient force to delaminate through approximately 10 $\mu$m of CNT forest growth (200 time steps). A critical detachment force of 3 nN resulted in 2927 unique CNTs experiencing sufficient force to delaminate. Assuming CNT detachment would occur when the critical force is exceeded, the simulated density after 10 $\mu$m of CNT forest growth would range from 4.3 - 5.1 x 10$^9$ CNT/cm$^2$ for 3 nN and 12 nN critical forces, respectively, in good agreement with the ESEM experimental observations shown in Figure \ref{fig:DelamSequence}b. For reference, 10 $\mu$m of CNT growth corresponds to approximately 200 seconds in Figure \ref{fig:DelamSequence}b. The consistency between mechanical simulation and experiment points to the reaction forces generated during CNT self-assembly as a significant driver of CNT for population decrease.

\subsection{CNT Forest Self-Termination}
The self-termination of CNT forest growth typically occurred after approximately 20-30 minutes of synthesis. During self-termination, an increasing fraction of CNTs rapidly cease growing, and the CNT forest height becomes static. The abrupt catalytic self-exhaustion of CNT forests has been experimentally measured by others using laser reflection measurements \cite{abruptTermination} or optical imagery \cite{roboFurnace}. In-situ TEM measurements of CNT forest self-termination are impractical because the CNTs themselves obscure the observation. Using in-situ ESEM synthesis, we observed that the activity of CNTs terminates in a distributed manner. The vertical growth of actively growing CNTs is resisted by CNTs that have stopped growing and are tethered to the substrate. Actively growing CNTs bend and buckle because of mechanical resistance, while the inactive, tethered CNTs experience a corresponding tensile force. The CNTs near the substrate exhibit a highly tortuous morphology because of this mechanical competition. The collective tensile load experienced by the inactive CNTs during this phase does not induce large-scale CNT delamination, indicating the load generated by active CNTs is well distributed among the inactive CNTs. The compressive load experienced by actively growing CNTs will transmit to the catalyst particle, where mechanochemical coupling may act to slow or deactivate catalyst activity \cite{dee2018situ}. 

\begin{figure}[h]
  \includegraphics[width=100mm]{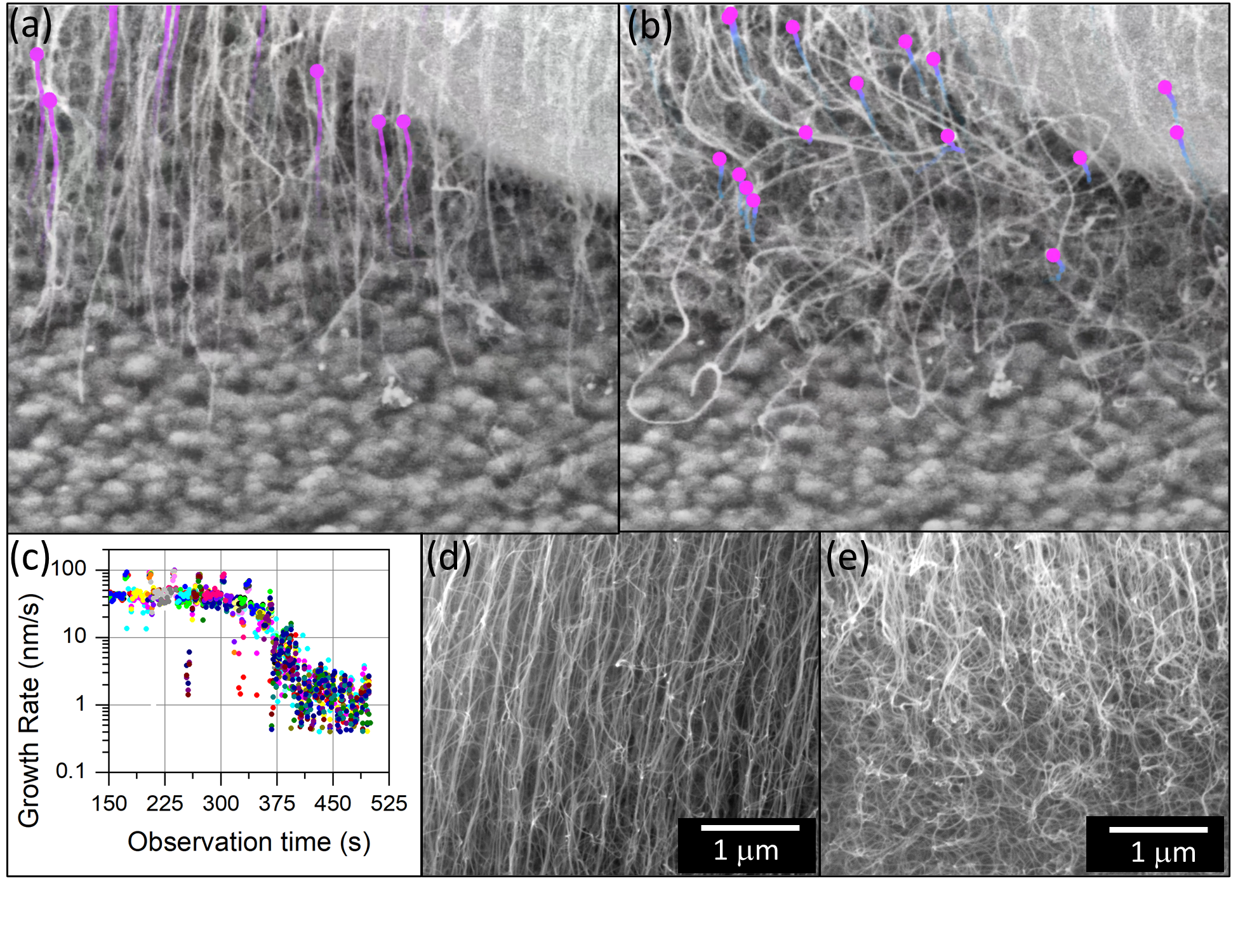}
  \centering
\caption{In-situ ESEM images of CNT forest self termination. The motion path of CNT features were measured using the Meta CoTracker algorithm beginning \textbf{a} 150 second after CNT forest synthesis initiation to \textbf{b} self termination. The solid circle indicates the location of an evaluation point in the current frame, while the trailing tail represents a time history of the point. \textbf{c} The vertical growth rate is nearly constant between 40-50 nm/s at the beginning of the analysis, followed by a growth rate decay beginning at 300 s and an increased growth rate disparity between CNTs. SEM images obtained after self termination show general CNT alignment near the top of the forest, \textbf{d}, and significant disorder at the bottom of the forest, \textbf{e}, where the CNT population experienced self-termination. }
  \label{fig:termination}
\end{figure}

The growth rate of CNTs before and during self-termination was evaluated using the Meta CoTracker algorithm, a transformer network that models the correlation of different points in time via specialised attention layers \cite{karaev2023cotracker}. By tracking points identical points across an image sequence, the motion path of CNTs was determined. Near the initiation of CNT forest growth (Figure \ref{fig:termination}a), the CNT path was largely vertical, normal to the growth substrate, and CNTs exhibited a high degree of alignment. The time history of evaluation points is represented by the tail of each point in Figure \ref{fig:termination}a-b. During self termination (Figure \ref{fig:termination}b), the growth of CNTs rapidly decelerated until most CNTs ceased growth, as observed by the shortened tails of evaluation points provided by CoTracker. The remaining active CNTs moved in a highly tortuous path because they were unable to advance in the vertical direction against the resistance of inactive CNTs. The growth rate observation began 150 seconds after the initiation of growth. The initial vertical growth rate 40-50 nm/sec is consistent with previous measurements obtained by DIC and presented in Figure \ref{fig:DIC}. This growth rate was nearly constant for the observation time between 150-300 seconds. The CNT growth rate rapidly decreased thereafter, until ceasing at approximately 450 seconds. The relative disparity in growth rates also increased during self-termination, indicating that growth rates of CNTs within the population were decelerating at different rates. While the CoTracker algorithm readily tracked the motion path of slowly terminating CNTs, it could not track the rapid and unpredictable tortuous path of active, rapidly growing CNTs at this phase; therefore, the most rapidly growing CNTs are not well reflected in the growth rate data in Figure \ref{fig:termination}c. The difference in CNT forest tortuosity is readily observed when comparing images obtained near the top of the forest (Figure\ref{fig:termination}d), which grew when CNTs were growing at a more uniform rate, to the morphology at the bottom (Figure \ref{fig:termination}e) which underwent self termination and a large disparity of CNT growth rates. A video of the CoTracker image sequence is found in supplementary information video S5.   
 
%\FloatBarrier

\section{Conclusion}
The evolving stages of CNT forest growth, including nucleation, self-assembly, delamination, and self-termination were directly observed using in-situ ESEM synthesis. The lifecycle of CNT forest synthesis is a dynamic process that undergoes significant evolution over time. We found that adventitious carbon selectively reduced and activated the underlying iron catalyst film in our in-situ experiments. The spatial selectivity of the CNT forest synthesis has not been previously reported in analogous in-situ TEM syntheses. We hypothesize that the small and localized heating zone within a large, quiescent gas chamber presented by the ESEM may establish a gas stagnation point above the heated zone that limits the availability of hydrocarbon gas to the catalyst. The adventitious carbon accelerated catalyst reduction and activated a carbon sink that may disrupt the native hydrodynamic stagnation point. 

CNT-CNT contacts were frequently established within 1 $\mu$m of the substrate, and those contacts persisted through the growth.  In-situ DIC measurements showed that the internal strain within a growing CNT forest volume is invariant as it lifts away from the growth substrate. Periodic lateral oscillation of growing CNT pillars are likely the result of collective forces established rapidly growing CNTs pushing against the rigid forest established above. These observations point to the critical significance of the near-substrate region in orchestrating the CNT forest morphology.  

Throughout coordinated CNT forest growth, mechanical reaction forces are generated between contacting CNTs that produce angular rocking at the CNT-substrate interface and lead to CNT detachment. The rotation of CNTs in response to reaction forces implies that iron catalyst nanoparticles are not static and must accommodate rotation at either the CNT-catalyst interface or the catalyst-substrate interface. To accommodate the CNT rotation, nanoparticles can reconfigure themselves while maintaining their catalytic activity, as has been observed by in-situ TEM synthesis \cite{huang}. Although the SEM alone cannot resolve the mechanisms behind the significant rotations of the CNTs around the catalyst nanoparticle, CNT rotation is a key mechanical boundary condition for CNT forest self-assembly and may hold kinetic implications relative to mechano-chemical feedback and the production of wall defects.

The tensile loading generated between interacting CNTs and transmitted to catalyst nanoparticles is likely a significant contributor to CNT delamination. CNTs that grow faster than the population average bend and deform after contacting slower growing CNTs. A tensile force is transmitted to the slower growing CNT. Delamination of slower growing CNTs occurs when sufficient mechanical energy is provided to overcome the adhesion energy of the CNT and the substrate. We observed two different types CNT delamination events. In some cases, the catalyst particle remained on the substrate, while in other cases the catalyst particle was removed from the substrate.  Simulations indicated that the tensile force generated from CNT-CNT interactions during synthesis could exceed several hundred nano-Newtons if CNTs were rigidly anchored to the substrate, far greater than loads observed during the delamination of CNT forests after synthesis. Because the reaction forces exceed the adhesion strength of the CNTs, delamination occurs and CNT density decreases with time. We measured a continuously decreasing CNT forest density that is consistent with previous SAXS measurements.

CNT forest self-termination occurred as progressively more CNTs terminated growth. After a relatively abrupt cessation of CNT forest height increase, a population of actively growing CNTs continued to lengthen and formed a highly tortuous layer near the growth substrate. The terminated CNTs anchored to the substrate acted as tethers to restrict further lengthening of the active CNTs. Evidence of mechanical competition is noticeable after collective forest termination in the form of a disorganized, wavy morphology of the CNT forest near the growth substrate (Figure \ref{fig:termination}). The dynamic observation of CNT forest self-termination provides context to the tortuous morphology frequently observed at the bottom of self-terminated CNT forests as quantified by previous Hermans Orientation Factor measurements made by SAXS.

The application of in-situ environmental SEM for chemical vapor deposition material synthesis fills a technological gap between in-situ TEM syntheses and other indirect in-situ measurements such as SAXS, Raman spectroscopy, and laser reflectivity. The technique provides a direct view of CNT forest assembly processes including CNT-CNT contact establishment, CNT delamination, CNT rotation, internal forest strain measurements, and forest growth rates. These measurements are critical to generate a high-fidelity model of CNT forest interactions and the process-structure-property relationships that are difficult to assess after forest syntheses. The integration of in-situ SEM and TEM observations to form a more comprehensive understanding of the interconnected relationships between detailed catalytic mechanisms and the assembly of CNT populations. Coupled with dynamic simulations, we foresee significant potential advances in understanding and controlling the synthesis of CNT forests.

\section{Experimental Section}

\subsection{In-Situ ESEM Substrate Preparation}
Prior to CNT forest synthesis, a Protochips E-chip was plasma treated using glow discharge at a pressure of 25-26 mBar with a current of 15 $\mu$A for 5 minutes. The catalyst film catalyst stack comprised of 10 nm aluminium oxide and 2 nm iron by ion-beam sputtering (South Bay Technologies IBS/e) at a base pressure of 1 x 10$^{-7}$ Torr.

\subsection{In-Situ Environmental Scanning Electron Microscope CNT Synthesis}
The heatign substrate was mounted in a Protochips Fusion 350 SEM stage operated within an FEI Quanta 600F environmental SEM. The MEMS-based heating substrates used in SEM experiments are identical to those used for TEM-based in-situ experiments. The substrate generates a uniform temperature region 50 x 50 $\mu$m and exceeds the 3 x 3 array of holes etched into the heated zone (see Figure \ref{fig:fig1}). The ESEM chamber with the substrate was pumped overnight (approximately 15 hours) to achieve a chamber pressure of approximately 8.5 x 10$^{-5}$ Pa. A cold finger was used to capture volatile impurities and water vapor. The cold finger tank was filled with a slurry of dry ice and acetone for one hour prior to in-situ testing. Liquid nitrogen was not used in the cold finger because of its low boiling temperature of -195.8 \textdegree C, which would freeze acetylene (−80.8 \textdegree C) on contact. Dry ice sublimates at a temperature of -78.5 \textdegree C, just above the freezing temperature of acetylene. 

\subsection{SEM Carbon Deposition}
Controlled exposure of the catalyst film to carbon increases CNT yield by accelerating catalyst nanoparticle reduction and dewetting \cite{dee}. In our experiments, adventitious carbon was deposited by electron beam induced deposition  \cite{van2008critical}. Hydrocarbon species on the sample surface or in the SEM chamber interact with the electron beam, the species crosslink and deposit as "diamond-like" carbon on the substrate\cite{miura}. This carbon layer is common in SEM and is visible as a black box \cite{soong}.

\subsection{CNT Forest Synthesis Simulation}
A 3D mechanical finite-element simulation was modeled the interactions and forces generated during CNT forest growth and self-assembly, analogous to a 2D simulation previously reported \cite{hajilounezhad2019evaluating, hajilounezhad2021predicting,maschmann2015integrated}. In summary, the simulation treats CNTs as numerous interconnected frame elements, with evaluation nodes located at the terminal end of each element. A new element was added to the base of each CNT at discrete time steps to simulate CNT growth. CNT-CNT van der Waals interactions were modeled as linear-elastic spring elements. A global stiffness matrix matrix consisting of all CNT elements and van der Waals elements was calculated at each time step prior to the addition of new CNT elements, and the displacement of all other nodes in the system is computed using the equation 

\begin{equation}
{F_t}=[K_t]{U_t}-[K_{t-1}]{U_{t-1}}
\end{equation}

where ${F_t}$ is the external force vector, $[K]$ is the global stiffness matrix, ${U}$ is the displacement vector, the subscript ${t}$ refers to the current time, and the subscript ${t-1}$ refers to the previous time step. This matrix equation was solved for each time step to compute the displacement vector, $U_t$, for all nodes in the system.

\subsection{Digital Image Correlation}
VIC-2D version 7 (Correlated Solutions, Inc.) software was used to evaluate the translation and strain of growing CNT forest pillars. The DIC evaluation criteria include 25 to 50-pixel evaluation subsets at 3–5 pixel steps. The low-pass filter was used to minimize the influence of image noise. The correlation algorithm utilized a 4-tap spline interpolation. A normalized squared differences method was used to accommodate potential changes in gray scale from image to image.

\subsection{Meta CoTracker}
The CoTracker algorithm \cite{karaev2023cotracker} was implemented using Python 3.11 running on a NVIDIA Quadro RTX 8000 GPU. A total of 62 evaluation points were tracked throughout a sequence of 350 in-situ ESEM images. Evaluation points were within the field of view for approximately 10-15 frames, at which time the evaluation point no longer contributed data.

\medskip
\textbf{Supporting Information} \par %Please delete the Suppporting Information statement if it is not applicable. Please supply Supporting Information in another file. Supporting information should not be provided in .tex format
Supporting Information is available from the Wiley Online Library or from the author.

% Acknowledgements
\medskip
\textbf{Acknowledgements} \par %delete if not applicable))
M.M., G.K., and T.H. acknowledge funding from NSF grant 1651538. M.M. and R.S. acknowledge funding from NSF grant 2026847. All authors acknowledge electron microscopy support from Dr. Dave Stalla and the MU Electron Microscopy Core.

\bibliography{CNT}
\bibliographystyle{Science}
%\textbf{References}\\

\end{document}